\overfullrule=0pt
\input harvmac
\lref\jusinskas{
   N.~Berkovits and R.~Lipinski Jusinskas,
  ``Light-Cone Analysis of the Pure Spinor Formalism for the Superstring,''
JHEP {\bf 1408}, 102 (2014).
[arXiv:1406.2290 [hep-th]].
}

\lref\brink{
 L.~Brink, M.~B.~Green and J.~H.~Schwarz,
  ``Ten-dimensional Supersymmetric {Yang-Mills} Theory With SO(8) - Covariant Light Cone Superfields,''
Nucl.\ Phys.\ B {\bf 223}, 125 (1983)..
}

\lref\BerkovitsYR{
  N.~Berkovits and O.~Chandia,
  ``Superstring vertex operators in an AdS(5) x S**5 background'',
Nucl.\ Phys.\ B {\bf 596}, 185 (2001).
[hep-th/0009168].
}

\lref\BerkovitsRB{
  N.~Berkovits,
``Covariant quantization of the superparticle using pure spinors,''
JHEP {\bf 0109}, 016 (2001).
[hep-th/0105050].
}

\lref\GalperinAV{
  A.~Galperin, E.~Ivanov, S.~Kalitsyn, V.~Ogievetsky and E.~Sokatchev,
``Unconstrained N=2 Matter, Yang-Mills and Supergravity Theories in Harmonic Superspace,''
Class.\ Quant.\ Grav.\  {\bf 1}, 469 (1984)..
}

\lref\vallilo{
  B.~Vallilo and L.~Mazzucato,
  ``The Konishi Multilpet at Strong Coupling,''
JHEP {\bf 1112}, 029 (2011).
[arXiv:1102.1219 [hep-th]].
}

\lref\MikhailovAF{
  A.~Mikhailov,
 ``Finite dimensional vertex,''
JHEP {\bf 1112}, 005 (2011).
[arXiv:1105.2231 [hep-th]].
}

\lref\mikh{
  A.~Mikhailov and R.~Xu, ``BRST cohomology of the sum of two pure spinors,''
to appear.
}

\lref\minahan{
  J.~Minahan,
  ``Holographic three-point functions for short operators,''
JHEP {\bf 1207}, 187 (2012).
[arXiv:1206.3129 [hep-th]].
}

\lref\BerkovitsGA{
  N.~Berkovits,
  ``Simplifying and Extending the AdS(5) x S**5 Pure Spinor Formalism,''
JHEP {\bf 0909}, 051 (2009).
[arXiv:0812.5074 [hep-th]].}



\lref\BerkovitsBT{
  N.~Berkovits,
  ``Pure spinor formalism as an N=2 topological string'',
JHEP {\bf 0510}, 089 (2005).
[hep-th/0509120].}



\lref\BerkovitsFE{
  N.~Berkovits,
  ``Super Poincare covariant quantization of the superstring'',
JHEP {\bf 0004}, 018 (2000).
[hep-th/0001035].
}


\lref\MazzucatoJT{
  L.~Mazzucato,
  ``Superstrings in AdS'',
[arXiv:1104.2604 [hep-th]].
}


\lref\HeslopNP{
  P.~Heslop and P.~S.~Howe,
  ``Chiral superfields in IIB supergravity'',
Phys.\ Lett.\ B {\bf 502}, 259 (2001).
[hep-th/0008047].
}







\lref\SohniusWK{
  M.~F.~Sohnius,
  ``Bianchi Identities for Supersymmetric Gauge Theories,''
Nucl.\ Phys.\ B {\bf 136}, 461 (1978).
}

\lref\ArutyunovGA{
  G.~Arutyunov and S.~Frolov,
  ``Foundations of the $AdS_5 \, x \, S^5$ Superstring. Part I,''
J.\ Phys.\ A {\bf 42}, 254003 (2009).
[arXiv:0901.4937 [hep-th]].
}

\lref\MetsaevIT{
  R.~R.~Metsaev and A.~A.~Tseytlin,
  ``Type IIB superstring action in AdS(5) x S**5 background,''
Nucl.\ Phys.\ B {\bf 533}, 109 (1998).
[hep-th/9805028].
}



\lref\HoweSRA{
  P.~S.~Howe and P.~C.~West,
  ``The Complete N=2, D=10 Supergravity'',
Nucl.\ Phys.\ B {\bf 238}, 181 (1984).
}

\def\bar{\overline}

\def\ads{AdS_5\times S^5}
\def\a{{\alpha}}

\def\ah{{\hat \a}}

\def\lh{{\widehat \lambda}}

\def\kb{{\bar \kappa}}

\def\l{{\lambda}}
\def\lb{{\overline\lambda}}
\def\lbh{{\widehat{\overline\lambda}}}

\def\lb{{\overline\lambda}}
\def\b{{\beta}}
\def\bh{{\hat\beta}}

\def\g{{\gamma}}

\def\d{{\delta}}
\def\e{{\epsilon}}

\def\xwt{{\widetilde x}}

\def\k{{\kappa}}
\def\kb{{\bar\kappa}}
\def\N{{\nabla}}

\def\L{{\Lambda}}
\def\Lh{{\widehat\Lambda}}
\def\lbh{{\widehat{\bar\l}}}
\def\O{{\Omega}}

\def\half{{1\over 2}}
\def\p{{\partial}}
\def\ads{{{$AdS_5\times S^5$}}}

\def\t{{\theta}}

\def\T{{\Theta}}
\def\Th{{\widehat\Theta}}

\def\th{{\widehat\theta}}

\def\lb{{\overline{\lambda}}}

\Title{\vbox{\baselineskip8pt
\hbox{}}}
{{\vbox{\centerline{Untwisting the Pure Spinor Formalism  }
\bigskip
\centerline{to the RNS and Twistor String}
\bigskip
\centerline{ in a Flat and $AdS_5\times S^5$ Background}}} }
\bigskip
\centerline{Nathan Berkovits}
\bigskip
\centerline{\it ICTP South American Institute for Fundamental Research}
\centerline{\it Instituto de F\'\i sica Te\'orica, UNESP - Univ. 
Estadual Paulista }
\centerline{\it Rua Dr. Bento T. Ferraz 271, 01140-070, S\~ao Paulo, SP, Brasil}

\bigskip

The pure spinor formalism for the superstring can be formulated as a twisted N=2 worldsheet theory with fermionic generators $j_{BRST}$ and composite $b$ ghost. After untwisting the formalism  to an N=1 worldsheet theory with fermionic stress tensor $j_{BRST}+b$, the worldsheet variables combine into N=1 worldsheet superfields $X^m$ and $\Theta^\alpha$ together with a superfield constraint relating $DX^m$ and $D\Theta^\alpha$. The constraint implies that the worldsheet superpartner of $\theta^\alpha$ is a bosonic twistor variable, and different solutions of the constraint give rise to the pure spinor or extended RNS formalisms, as well as a new twistor-string formalism with manifest N=1 worldsheet supersymmetry.

These N=1 worldsheet methods generalize in curved Ramond-Ramond backgrounds, and a manifestly N=1 worldsheet supersymmetric action is proposed for the superstring in an $AdS_5\times S^5$ background in terms of the twistor superfields. This $AdS_5\times S^5$ worldsheet action is a remarkably simple fermionic coset model with manifest $PSU(2,2|4)$ symmetry and might be useful for computing $AdS_5\times S^5$ superstring scattering amplitudes.

\bigskip
\centerline{\it Dedicated to the memory of Mario Tonin}

\vskip .2in

\Date {April 2016}
\newsec{Introduction}

The pure spinor formalism for the superstring \ref\BerkovitsFE{
  N.~Berkovits,
  ``Super Poincare covariant quantization of the superstring,''
JHEP {\bf 0004}, 018 (2000).
[hep-th/0001035].} has the advantage over the Ramond-Neveu-Schwarz (RNS) formalism in that is manifestly spacetime supersymmetric. This simplifies the computation of
multiloop superstring amplitudes since there is no sum over spin structures, and allows the description of Ramond-Ramond superstring backgrounds such as $AdS_5\times S^5$. However, the
pure spinor formalism has the disadvantage that it is not manifestly worldsheet supersymmetric. This complicates the construction of the $b$ ghost and integrated vertex operators, and introduces subtleties associated with regulators \ref\BerkovitsVI{
  N.~Berkovits and N.~Nekrasov,
  ``Multiloop superstring amplitudes from non-minimal pure spinor formalism,''
JHEP {\bf 0612}, 029 (2006).
[hep-th/0609012].
} and contact terms \ref\BerkovitsRPA{
  N.~Berkovits and E.~Witten,
  ``Supersymmetry Breaking Effects using the Pure Spinor Formalism of the Superstring,''
JHEP {\bf 1406}, 127 (2014).
[arXiv:1404.5346 [hep-th]].
} needed to preserve BRST invariance.

Although the pure spinor formalism is not manifestly worldsheet supersymmetric, it has a twisted
N=2 worldsheet supersymmetry in which the two fermionic N=2 generators are the BRST current and the $b$ ghost \ref\BerkovitsBT{
  N.~Berkovits,
  ``Pure spinor formalism as an N=2 topological string,''
JHEP {\bf 0510}, 089 (2005).
[hep-th/0509120].
}. In this paper, the N=2 worldsheet supersymmetry will be untwisted and the
pure spinor formalism will be described in a manifestly N=1 worldsheet supersymmetric and
d=10 spacetime supersymmetric manner in terms of the N=1 worldsheet superfields
\eqn\wss{X^m = x^m + \k \psi^m, \quad \T^\a = \t^\a + \k \L^\a, \quad
\Phi_\a = \O_\a + \k h_\a,}
where $\k$ is the anticommuting coordinate, $(x^m, \t^\a)$ are the usual d=10
superspace variables, $(\psi^m, \L^\a)$ are their worldsheet superpartners, and 
$(\O_\a, h_\a)$ are the conjugate momenta to $(\L^\a, \t^\a)$. 

The N=1 worldsheet superfields of \wss\ are constrained to satisfy
\eqn\consw{\lb\g^m \Phi =0, \quad (\lb\g^m)_\a (DX_m -\half D\T\g_m\T) =0,}
where $\lb_\a$ is a fixed d=10 pure spinor satisfying $\lb\g^m\lb=0$. Although
the constraints of \consw\ break manifest Lorentz covariance, one can solve these
constraints using three different methods to produce three different Lorentz-covariant
descriptions of the superstring. 

The first method is to solve for $\psi^m$ and $\L^\a$ in terms of $(x^m, \t^\a, \l^\a)$
where $\l^\a$ is a d=10 pure spinor satisfying $\l\g^m\l=0$. This method produces the
pure spinor formalism which is manifestly spacetime supersymmetric but not manifestly
worldsheet supersymmetric, and where the N=1 fermionic generator is the sum of the pure
spinor BRST current and $b$ ghost. 

The second method is to solve for $\t^\a$ and $\L^\a$ in terms of $(x^m, \psi^m, \t'^\a, \l^\a)$
where $\t'^\a$ is constrained to satisfy $\t' \g^m\l=0$ and $\l^\a$ is constrained to
satisfy $\l\g^m\l=0$. This method is manifestly worldsheet
supersymmetric where $\l^\a$ is the worldsheet superpartner of $\t'^\a$, but is not manifestly
spacetime supersymmetric. One can argue that $(\t'^\a, \l^\a)$ decouples from physical vertex
operators and scattering amplitudes, so this method produces an ``extended'' version
of the RNS formalism where $X^m = x^m+\k\psi^m$ plays the role of the usual RNS matter superfield.

Finally, the third method is to solve for $x^m$ and $\psi^m$ in terms of
$(\L^\a, \t^\a)$ and its conjugate momenta $(\O_\a, h_\a)$. 
This method preserves both manifest worldsheet supersymmetry and spacetime
supersymmetry, and produces a twistor description of the superstring in which 
$(\L^\a, \O_\a)$ are d=10 twistor variables which replace the $x^m$ spacetime variable. 

There are several similarities of this worldsheet supersymmetric twistor description with earlier twistor descriptions of the superstring in \ref\WittenNT{
  E.~Witten,
  ``Twistor - Like Transform in Ten-Dimensions,''
Nucl.\ Phys.\ B {\bf 266}, 245 (1986).
}\ref\SorokinNJ{
  D.~P.~Sorokin, V.~I.~Tkach, D.~V.~Volkov and A.~A.~Zheltukhin,
  ``From the Superparticle Siegel Symmetry to the Spinning Particle Proper Time Supersymmetry,''
Phys.\ Lett.\ B {\bf 216}, 302 (1989)\semi
  M.~Tonin,
  ``World sheet supersymmetric formulations of Green-Schwarz superstrings,''
Phys.\ Lett.\ B {\bf 266}, 312 (1991)\semi

  N.~Berkovits,
  ``The Heterotic Green-Schwarz superstring on an N=(2,0) superworldsheet,''
Nucl.\ Phys.\ B {\bf 379}, 96 (1992).
[hep-th/9201004]\semi

  F.~Delduc, A.~Galperin, P.~S.~Howe and E.~Sokatchev,
  ``A Twistor formulation of the heterotic D = 10 superstring with manifest (8,0) world sheet supersymmetry,''
Phys.\ Rev.\ D {\bf 47}, 578 (1993).
[hep-th/9207050]\semi

  A.~Galperin and E.~Sokatchev,
  ``A Twistor formulation of the nonheterotic superstring with manifest world sheet supersymmetry,''
Phys.\ Rev.\ D {\bf 48}, 4810 (1993).
[hep-th/9304046]\semi

  D.~P.~Sorokin,
  ``Superbranes and superembeddings,''
Phys.\ Rept.\  {\bf 329}, 1 (2000).
[hep-th/9906142]\semi

  M.~Matone, L.~Mazzucato, I.~Oda, D.~Sorokin and M.~Tonin,
  ``The Superembedding origin of the Berkovits pure spinor covariant quantization of superstrings,''
Nucl.\ Phys.\ B {\bf 639}, 182 (2002).
[hep-th/0206104].

}, however, these earlier twistor descriptions were mostly for the heterotic superstring
whereas this twistor description is only for the Type II superstring. It would be very interesting
to study the relation of these twistor descriptions to each other, as well as to the more recent
twistor superstrings which describe either N=4 d=4 super-Yang-Mills \ref\WittenNN{
  E.~Witten,
  ``Perturbative gauge theory as a string theory in twistor space,''
Commun.\ Math.\ Phys.\  {\bf 252}, 189 (2004).
[hep-th/0312171]\semi 
  N.~Berkovits,
  ``An Alternative string theory in twistor space for N=4 superYang-Mills,''
Phys.\ Rev.\ Lett.\  {\bf 93}, 011601 (2004).
[hep-th/0402045].
} or d=10 supergravity \ref\MasonSVA{
  L.~Mason and D.~Skinner,
  ``Ambitwistor strings and the scattering equations,''
JHEP {\bf 1407}, 048 (2014).
[arXiv:1311.2564 [hep-th]]\semi
N.~Berkovits,
  ``Infinite Tension Limit of the Pure Spinor Superstring,''
JHEP {\bf 1403}, 017 (2014).
[arXiv:1311.4156 [hep-th]].
}.

In a flat background, the N=(1,1) worldsheet supersymmetric action for the Type II twistor
superstring is
\eqn\tts{S = \int d^2 z d^2 \k [-\Phi_\a \bar D \T^\a + \widehat\Phi_\ah D \Th^\ah
-{1\over 8} (\T\g^m D\T)(\Th\g_m \bar D\Th) +{1\over 8}(\Th\g^m D\Th)(\T\g_m\bar D\T)]}
where $D={\p\over\p\k} + \k\p_z$ and $\bar D= {\p\over\p\kb} + \kb\bar\p_{\bar z}$,  $(\T^\a, \Phi_\a, \Th^\ah, \widehat\Phi_\ah)$ are N=(1,1) worldsheet
superfields and $\a,\ah=1$ to 16 are d=10 spinor indices of the same/opposite chirality
for the Type IIB/IIA superstring. This action is manifestly invariant under both N=(1,1) worldsheet
supersymmetry and d=10 N=2 spacetime supersymmetry
which transforms the worldsheet superfields as
\eqn\uus{\d\T^\a = \e^\a, \quad \d\Th^\ah = \widehat\e^\ah,}
$$\d\Phi_\a = {1\over 4}(\e\g^m\T + \widehat\e\g^m\Th) (\g_m D\T), \quad
\d\widehat\Phi_\ah ={1\over 4} (\e \g^m\T + \widehat\e\g^m\Th) (\g_m\bar D\Th).$$

Surprisingly, when expressed in terms of these twistor superfields, the Type IIB superstring action
in an $AdS_5\times S^5$ background
takes the extremely simple form 
\eqn\ttads{S =r^2 \int d^2 z d^2 \k [ D \T^J_R \bar D \tilde\T_J^R + D\T^J_S \tilde\T^S_K \bar D\T^K_R \tilde\T^R_J]}
where $r$ is the $AdS$ radius, $R=1$ to 4 are $SO(4,2)$ spinor indices for $AdS_5$, $J =1$ to 4 are
$SO(6)$ spinor indices for $S^5$, the 16 components of the superfield $\T^J_R$ are obtained by decomposing $\T^\a + i \Th^\ah$ under $SO(4,2)\times SO(6)$, and the 16 components of the superfield $\tilde\T_J^R$ are obtained by decomposing
$\T^\a - i\Th^\ah$. This $AdS_5\times S^5$ twistor-string action is manifestly invariant under both N=(1,1) worldsheet supersymmetry
and under $PSU(2,2|4)$ where the 32 spacetime supersymmetries transform the
worldsheet superfields as
\eqn\susytr{\d\T_R^J = \e_R^J + \T^J_S \tilde\e^S_K \T^K_R, \quad 
\d\tilde\T_J^R = \tilde\e^R_J - \tilde\e^R_K \T^K_S \tilde\T^S_J - \tilde\T^R_K \T^K_S \tilde\e^S_J.}

Hopefully, the simple form of \ttads\ will be useful for constructing vertex operators and computing
superstring scattering amplitudes in an $AdS_5\times S^5$ background. 
But before 
constructing vertex operators and computing scattering amplitudes using this
twistor-string action in an $AdS\times S^5$ background, it will be necessary to
better understand the vertex operators and scattering amplitudes using the
twistor-string action in a flat background of \tts.

In section 2.1, the N=2 worldsheet supersymmetry of the pure spinor formalism is untwisted to an N=1 worldsheet supersymmetry
with the fermionic generator $G = j_{BRST}+b$, and the constrained N=1 worldsheet superfields 
$[X^m, \T^\a, \Phi_\a]$ satisfying \consw\ are defined. In section 2.2, the N=1 worldsheet supersymmetric action with manifest d=10 supersymmetry is
constructed for the superstring in a flat background in terms of these constrained superfields. In section 2.3, the U(1) generator $J = -\l^\a w_\a$ is used to define physical states whose integrated vertex operator is required to be N=1
superconformally invariant and have zero or negative U(1) charge. In section 2.4, physical vertex operators are constructed for massless states and the Siegel gauge-fixing condition $b_0=0$ is clarified. In section 2.5, a new tree amplitude prescription is given for the pure spinor formalism based on the
untwisted approach which matches the RNS tree amplitude prescription in the ${\cal F}_1$ picture.
In section 2.6, an alternative solution to the superfield constraints of \consw\ is shown to produce an extended version of the RNS formalism where the $[\T^\a, \Phi_\a]$ superfields decouple from the
$X^m$ superfield. And in section 2.7, the N=1 worldsheet supersymmetric approach to
the pure spinor formalism is generalized to curved heterotic and Type II supergravity backgrounds.

In section 3.1, a third solution to the superfield constraints of \consw\ for the Type II superstring is described which replaces the usual spacetime variable $x^m$ with twistor variables and solves for
$X^m$ in terms of $[\T^\a, \Phi_\a]$. In section 3.2, N=1 worldsheet superconformal generators are constructed for this Type II twistor-string formalism and a U(1) generator corresponding to the projective weight of d=10 twistors is used to define physical states. In section 3.3, the N=(1,1) worldsheet supersymmetric twistor-string action in a flat background of \tts\ is constructed and shown to be equivalent to the usual pure spinor Type II superstring action
up to a BRST-trivial term. In section 3.4, this twistor-string action is
generalized in an $AdS_5\times S^5$ background to the remarkably simple action of \ttads\ which has manifest $PSU(2,2|4)$ symmetry
and reduces in the large radius limit to the action in a flat
background of \tts. Finally, in section 3.5, the $AdS_5\times S^5$ twisor-string action is
written in $SO(10,2)$ notation and a U(1) generator involving d=12 pure spinors is used to define physical states. 

\newsec{Untwisting the Pure Spinor Formalism}

\subsec{ N=1 generators and superfields}

In a flat background, the left-moving variables of the pure spinor formalism for the superstring are described in conformal gauge by the free worldsheet action
\eqn\purefree{S = \int d^2 z (\half \p x^m \bar \p x_m + p_\a \bar\p\t^\a + w_\a \bar\p \l^\a),}
where $(x^m, \t^\a)$ are the usual N=1 d=10 superspace variables for $m=0$ to 9 and $\a=1$ to 16, $p_\a$ is the conjugate momenta to $\t^\a$, 
$\l^\a$ is a d=10 pure spinor variable satisfying $\l\g^m\l=0$, and $w_\a$ is the conjugate momentum to $\l^\a$ which is defined up to the gauge transformation $\d w_\a = 
f^m (\g_m\l)_\a$.

As discussed in \BerkovitsBT, this pure spinor formalism can be interpreted as a
topologically twisted N=2 worldsheet superconformal field theory with fermionic left-moving
generators 
\eqn\ntwo{G^+ = j_{BRST} = \l^\a d_\a,}
$$G^- = b =  -w_\a \p\t^\a +{1\over{2(\l\lb)}} [(\pi^m (\lb\g_m d)  +(w\g_m\lb)(\l\g^m\p\t)]$$
where $\oint G^+$ is the BRST charge used to define physical states, $G^-$ is
the composite $b$ ghost used for computing loop amplitudes, $\lb_\a$ is a fixed pure spinor on a patch defined by $(\l^\a \lb_\a)\neq 0$, and $\pi^m$
and $d_\a$ are the spacetime supersymmetric operators
\eqn\susydef{\pi^m = \p x^m -\half \p\t\g^m \t, \quad d_\a = p_\a -\half  \p x_m (\g^m\t)_\a -{1\over 8} 
(\t\g^m\p\t)(\g_m\t)_\a.}
Using the OPE's from the free worldsheet action of \purefree,
one can verify that $G^+$ and $G^-$ are nilpotent operators satisfying the relation 
\eqn\stress{\{\oint G^+, G^-\} = T_{twisted} = -\half \p x^m \p x_m - p_\a \p\t^\a - w_\a \p\l^\a}
for any choice of $\lb_\a$. Although $G^-$ can be Lorentz-covariantized by treating
$\lb_\a$ as a non-minimal worldsheet variable, this non-minimal version of the pure spinor formalism will not be discussed here and $\lb_\a$ will be assumed to be fixed on each patch. Furthermore, we will be ignoring all normal-ordering terms and central charges thoughout this paper such as the term proportional to $\lb_\a\p^2\t^\a$ in $G^-$. Hopefully, the non-minimal formalism and normal-ordering contributions will be treated in a later paper.  

To untwist the N=2 generators of \ntwo, define the N=1 generator
\eqn\none{G = G^+ + G^- = \l^\a d_\a - w_\a \p\t^\a + {1\over{2(\l\lb)}} [(\pi^m (\lb\g_m d)  +(w\g_m\lb)(\l\g^m\p\t)]}
which satisfies the OPE of an N=1 superconformal stress tensor
\eqn\oneope{G(y) G(z) \to 2 (y-z)^{-1} T(z)  }
where
\eqn\unT{T = - \half \p x^m \p x_m - p_\a \p\t^\a - \half (w_\a \p\l^\a - \l^\a \p w_\a)}
is the untwisted stress tensor with $(\l^\a, w_\a)$ of $+\half$ conformal weight, and the central charge contribution in \oneope\ is
being ignored.

Under the N=1 worldsheet supersymmetry generated by $G$ of \none, the bosonic worldsheet
superpartner $G\t^\a$ of $\t^\a$ is 
\eqn\supertheta{\L^\a = \l^\a + {1\over{2(\l\lb)}} \pi^m (\g_m\lb)^\a} 
and the fermionic worldsheet superpartner $Gx^m$ of $x^m$ is 
\eqn\superx{\psi^m = \half\L\g^m\t - {1\over{2(\l\lb)}} (\lb\g^m d).}
So one can define N=1 worldsheet superfields
\eqn\superf{X^m = x^m + \k \psi^m, \quad \Theta^\a = \t^\a + \k \L^\a}
where $\k$ is an anticommuting parameter, which transform covariantly under N=1 worldsheet supersymmetry transformations and transform under the d=10
spacetime supersymmetry transformations as $\d\T^\a = \e^\a, \d X^m =-\half \e\g^m\T$.

Furthermore,
the conjugate momenta variables $p_\a$ and $w_\a$ can be combined into the worldsheet superfield
\eqn\superPhi{\Phi_\a =
  w_\a -  {1\over{2(\l\lb)}}(w\g_m\lb)(\g^m \l) + \k [ d_\a  - {1\over{2(\l\lb)}}(d\g_m\lb)(\g^m \l)_\a  -  {1\over{2(\l\lb)^2}} \lb_\a (\lb \g_m d)\pi^m] }
where $\pi^m$ and $d_\a$ are defined in \susydef. Note that $\Phi_\a$ has
conformal weight $+\half$ and transforms covariantly under N=1
worldsheet supersymmetry, and is spacetime supersymmetric.
The N=1 superfields $(X^m, \T^\a, \Phi
_\a)$ are not independent and satisfy the worldsheet and spacetime supersymmetric constraints
\eqn\consone{ (\g^m \lb)^\a (D X_m -\half D \T \g_m \T) =0, \quad (\lb \g^m)^\a \Phi_\a  =0}
where $D= {\p\over{\p\k}} + \k {\p\over{\p z}}$. 

As will be shown later, different solutions of the constraints of \consone\ will describe either the pure spinor formalism, an extended version of the RNS formalism, or a new twistor formalism of the superstring.

\subsec{ Worldsheet supersymmetric action}

To construct the N=(1,0) worldsheet supersymmetric action for the heterotic superstring, generalize the superfields of \superf\ and \superPhi\ to the off-shell N=(1,0) superfields 
\eqn\offs{X^m = x^m + \k \psi^m, \quad \T^\a = \t^\a + \k \L^\a, \quad \Phi_\a = \O_\a + \k h_\a,}
where $(x^m, \psi^m, \t^\a, \L^\a, \O_\a, h_\a)$ are treated as independent components.
For the heterotic superstring in a flat background, the N=(1,0) worldsheet action in terms of these superfields is
\eqn\het{S = \int d^2 z d\k [ \Phi_\a \bar\p\T^\a +\half\Pi_{\k}^m \bar\Pi_{\bar z m} +B^{het}_{\k \bar z}
+ (\lb\g^m L)\Pi_{\k m} + M^m (\lb\g_m\Phi)]}
where 
\eqn\Pis{\Pi_\k^m = DX^m -\half D\T\g^m \T, \quad
\Pi_z^m = \p X^m -\half \p\T\g^m \T, \quad \bar\Pi_{\bar z}^m = \bar\p X^m -\half \bar\p\T\g^m\T,}
\eqn\Bs{B^{het}_{\k \bar z} ={1\over 4}[ (D\T\g_m \T) \bar\p X^m - DX^m (\bar\p\T\g_m \T)], \quad
B^{het}_{z \bar z} = {1\over 4}[(\p\T\g_m \T) \bar\p X^m - \p X^m (\bar\p\T\g_m \T)],}
$B^{het}_{z\bar z} $ is the usual heterotic Green-Schwarz two-form, $B^{het}_{\k\bar z}$ is obtained from $B^{het}_{z\bar z}$ by replacing $\p_z$ with $D$, $L_\a$ and $M^m$ are Lagrange multiplier superfields enforcing the constraints of \consone, and the right-moving fermions of the heterotic superstring which generate the 
$SO(32)$ or $E_8\times E^8$ gauge groups will be ignored throughout this paper. 

Performing the Grassmann integral over $\k$ and imposing the constraints of 
\consone, the action of \het\ is equal to
\eqn\heta{S = \int d^2 z  [ D\Phi_\a \bar\p\T^\a + \Phi_\a \bar\p D\T^\a +\half(\Pi_{z}^m -\half D\T\g^m D\T)\bar\Pi_{\bar z m} -\half \Pi_{\k m} (\bar \p\Pi_\k^m + D\T\g^m \bar\p\T)}
$$ +B^{het}_{z \bar z} +\half (D\T\g^m \bar\p\T) \Pi_\k +{1\over 4}
 (D\T\g^m D\T) \bar\Pi_{\bar z} ]$$
$$ = \int d^2 z  [ D\Phi_\a \bar\p\T^\a + \Phi_\a \bar\p D\T^\a +\half\Pi_{z}^m \bar\Pi_{\bar z m} - \Pi_\k^m  D\T\g_m \bar\p\T
 +B^{het}_{z \bar z} ]$$
\eqn\hetlast{ = \int d^2 z [  h_\a \bar\p\t^\a + \O_\a\bar\p\L^\a  - (\psi^m -\half \L\g^m\t) (\L\g_m)_\a \bar\p \t^\a  + \half
\pi^m \bar\pi_m + B^{het}_{z\bar z} ]}
where $\O_\a$ and $h_\a$ are constrained to satisfy $\lb\g^m\O = \lb\g^m h=0$. Finally, one can define 
\eqn\deffO{\O_\a = w_\a -   {1\over{2(\l\lb)}}(w\g_m\lb)(\g^m \l), \quad
d_\a = h_\a - (\L\g_m)_\a (\psi^m -\half \L\g^m \t),}
to obtain the heterotic pure spinor action of \purefree
\eqn\hetp{S = \int d^2 z  [ d_\a \bar\p\t^\a + w_\a \bar\p\l^\a +\half \pi^m \bar\pi_m + B^{het}_{z\bar z}]}
$$=  \int d^2 z [\half \p x^m \bar \p x_m + p_\a \bar\p\t^\a + w_\a \bar\p \l^\a]$$ 
where the relation of $p_\a$ and $d_\a$ is defined in \susydef.


For the Type II superstring, one generalizes the N=(1,0) superfields of \offs\ to
N=(1,1) off-shell superfields
\eqn\supertwo{X^m =x^m + \k\psi^m + \bar\k\widehat\psi^m + \k\bar\k f^m,}
$$  \T^\a = \t^\a + \k\L^\a + \bar\k \rho^\a + \k\bar\k s^\a,\quad
 \Th^\ah =\th^\ah + \bar\k\widehat\L^\ah + \k \widehat\rho^\ah + \k\bar\k \widehat s^\ah,$$
 $$ \Phi_\a =\O_\a + \k h_\a + \bar\k r_\a + \k\bar\k\xi_\a , \quad
  \widehat\Phi_\ah =\widehat\O_\ah + \bar\k \widehat h_\ah + \k \widehat r_\ah + \k\bar\k\widehat\xi_\ah,$$ 
 where
$(z,\bar z, \k, \kb)$ are the parameters of N=(1,1) worldsheet superspace and $\a$ and $\ah$ denote spinors of the same/opposite chirality for the Type IIB/IIA superstring. In terms of these
N=(1,1) worldsheet superfields, the Type II worldsheet supersymmetric action in a flat
background is
\eqn\twoact{S = \int d^2 z d\k d\kb [ - \Phi_\a \bar D\T^\a + \widehat\Phi_\ah D\Th^\ah +\half \Pi_\k^m \bar\Pi_{\bar \k m}
+ B^{II}_{\k\bar\k}}
$$+ (\lb\g_m L)\Pi_k^m +M_m (\lb\g^m\Phi) 
+ (\lbh\g_m \widehat L)\bar \Pi_{\bar \k}^m
+\widehat M_m (\lbh\g^m\widehat\Phi)
],$$
where $[L_\a, M^m]$ and $[\widehat L_\ah, \widehat M^m]$ are
Lagrange multipliers for the left and right-moving constraints
\eqn\twocon{\Pi_\k^m (\g^m\lb)^\a =0, \quad \lb\g^m\Phi =0, \quad
\bar\Pi_\kb^m (\g^m\lbh)^\ah =0, \quad \lbh\g^m \widehat\Phi =0,}
$\lb_\a$ and
$\lbh_\ah$ are two fixed pure spinors satisfying $\lb_\a \l^\a \neq 0$ and
$\lbh_\ah \lh^\ah \neq 0$, 
$D= {\p\over{\p\k}} + \k {\p\over{\p z}}$ and 
$\bar D= {\p\over{\p\kb}} + \kb {\p\over{\p \bar z}}$, 
\eqn\defpitwo{
\Pi_\k^m = D X^m-\half D\Theta \gamma^m \T-\half D\Th\g^m \Th, \quad
\Pi_z^m = \p X^m -\half\p\Theta \gamma^m \T -\half\p\Th\g^m \Th, }
\eqn\pithree{\bar\Pi_{\bar\k}^m = \bar D X^m -\half\bar D\Theta \gamma^m \T-\half\bar D \Th\g^m \Th, \quad
\bar\Pi_{\bar z}^m = \bar\p X^m-\half \bar\p\Theta \gamma^m \T-\half\bar\p \Th\g^m \Th,}
\eqn\btwo{B^{II}_{\k\bar\k} =   {1\over 4}[(D\T\g_m \T -D \Th\g_m \Th) \bar  D X^m -
D X^m (\bar D\T\g_m \T - \bar D\Th\g_m \Th)    }
$$
-\half (D\T\g_m \T)(\bar D\Th\g^m \Th) +\half (D\Th\g^m  \Th)(\bar D\T\g_m \T)],$$
\eqn\bthree{
 B^{II}_{z\bar z} ={1\over 4}[   (\p\T\g_m \T - \p\Th\g_m \Th) \bar  \p X^m -
\p X^m (\bar\p\T\g_m \T - \bar\p\Th\g_m \Th)    }
$$
-\half (\p\T\g_m \T)(\bar\p\Th\g^m \Th) +\half (\p\Th\g^m  \Th)(\bar\p\T\g_m \T)],$$
and $B^{II}_{\k\kb}$ is the usual Type II Green-Schwarz two-form $B_{z\bar z}$ field with $\p\over{\p z}$ and $\p\over{\p\bar z}$ replaced by $D$ and $\bar D$.

After shifting $\Phi_\a$ and $\widehat\Phi_\ah$, integrating over $\k$ and $\bar \k$, and solving for
auxiliary fields, the action of \twoact\ reduces to
\eqn\acttype{S = \int d^2 z  [ h_\a \bar\p\t^\a + \O_\a \bar\p \L^\a +
\widehat h_\ah \p\th^\ah + \widehat\O_\ah \p \Lh^\ah -
 (\psi^m - \half\L\g^m\t) (\L\g_m)_\a \bar\p \t^\a }
 $$ - (\widehat\psi^m -\half \widehat\L\g^m\th) (\widehat\L\g_m)_\a \p \th^\a +
\half\pi^m \bar\pi_m + B^{II}_{z\bar z} ]$$
where 
$[\O_\a, \widehat\O_\ah, h_\a, \widehat h_\ah]$ are constrained to satisfy 
$\lb\g^m\O =\lbh\g^m \widehat\O =  \lb\g^m h=\lbh \g^m \widehat h=0$. Defining 
\eqn\deffOtwo{\O_\a = w_\a -   {1\over{2(\l\lb)}}(w\g_m\lb)(\g^m \l), \quad
d_\a = h_\a - (\L\g_m)_\a (\psi^m - \half\L\g^m \t),}
$$\widehat\O_\ah = \widehat w_\ah -   {1\over{2(\lh\lbh)}}(\widehat w\g_m\lbh)(\g^m \lh), \quad
\widehat d_\a = \widehat h_\ah -(\widehat\L\g_m)_\ah (\widehat\psi^m - \half\widehat\L\g^m \th), $$
one obtains the Type II pure spinor action
\eqn\purefreetwo{S = \int d^2 z [\half \pi^m \bar \pi_m + B^{II}_{z\bar z} + d_\a \bar\p\t^\a + w_\a \bar\p \l^\a + 
\widehat d_\ah \p \th^\ah + \widehat w_\ah \p \lh^\ah]}
$$= \int d^2 z [\half \p x^m \bar \p x_m + p_\a \bar\p\t^\a + w_\a \bar\p \l^\a + 
\widehat p_\ah \p \th^\ah + \widehat w_\ah \p \lh^\ah].$$

Although the manifestly worldsheet supersymmetric actions of \het\ and \twoact\ are not manifestly Lorentz-covariant because
of the presence of $\lb_\a$ in the constraints of \consone, one can solve these constraints to obtain the manifestly Lorentz-covariant action of the pure spinor formalism
which is, however, not manifestly worldsheet supersymmetric. As will be shown later, there
are alternative ways to solve the constraints of \consone\  which either lead to
the extended RNS formalism
or to the twistor string formalism. However,
before discussing the relation of \het\ and \twoact\ to the extended RNS and twistor-string formalisms,
it will be shown how to construct vertex operators and compute tree-level scattering
amplitudes using this N=1 worldsheet supersymmetric description of the pure spinor formalism.

\subsec{U(1) generator}

As expected for an N=1 superconformal field theory, physical vertex operators $V$ should be  N=1 superconformal primary fields of conformal weight $+\half$ so that the integrated vertex operator $\int GV= \int dz d\k V$ is N=1 superconformally invariant. But after imposing the constraints of \consone\ and fixing the N=1 superconformal invariance,
the superfields $[X^m, \T^\a, \Phi_\a]$ contain $30+30$ worldsheet
variables.  So one needs to impose additional requirements if one wants to reproduce the usual superstring spectrum depending
in light-cone gauge on only $8+8$ worldsheet variables. 

To obtain the additional requirements, consider the U(1) generator
\eqn\jgen{J =- \l^\a w_\a }
which has the OPE's 
\eqn\njg{ J(y) G(z) \to (y-z)^{-1} (G^+ - G^-)}
where $G = G^+ + G^-$ and $G^\pm$ are defined in \ntwo\ and carry $\pm 1$ U(1) charge
with respect to $\oint J$.
Since the integrated vertex operator $\int GV$ is N=1 superconformally invariant, it would be N=2 superconformally invariant if it had no poles with $J$ since this would imply that $\int GV$ has no poles with either $G^+$ or $G^-$. Although this condition on the vertex operator would be too restrictive, an appropriate condition is that
$\int GV$ must have only terms of zero or negative U(1) charge with respect to $\oint J$. Defining $\int (GV)_n$ to be the term in $\int GV$ with U(1) charge $n$, this condition combined with N=1 superconformal invariance implies that
$[\oint G^+, \int (GV)_0]=0$. 

It will later be shown that charge conservation implies that the terms with negative U(1) charge in $\int GV$ do not contribute to tree amplitudes. So at least for tree amplitudes, the integrated vertex operator can be identified with $\int (GV)_0$ which is annihilated by
$\oint G^+$. Furthermore, it will be required that the integrated vertex operator of zero U(1) charge,
$\int (GV)_0$, is independent of the fixed pure spinor $\lb_\a$ and is therefore globally
defined on the pure spinor space. So in addition to requring that
$\int GV$ is N=1 superconformally invariant, it will also be required that $\int (GV)_n=0$ for
$n$ positive and that $\int (GV)_0$ is globally defined on the pure spinor space, i.e.
$\int (GV)_0$ is independent of $\lb_\a$ and is invariant under the gauge transformation $\d w_\a = f^m (\g_m\l)_\a$. By fixing the way that the vertex operator depends on 11 components of $\l^\a$ and $\t^\a$ and their conjugate momenta, these additional
requirements will reduce the degrees of freedom in physical vertex operators from $30+30$ worldsheet variables to $8+8$ worldsheet variables.

\subsec{Massless vertex operators}

N=1 superconformal invariance implies that the open superstring unintegrated
massless vertex operator of conformal weight $+\half$ has the form
\eqn\openv{ V =  D\T^\a A_\a (X, \T)+  \Pi_\k^m A_m (X, \T) +  \Phi_\a W^\a(X, \T)}
where $(A_\a, A_m, W^\a)$ are spacetime superfields with momentum $k^m$ satisfying $k^m k_m =0$. By acting on $V$ with the worldsheet superspace derivative $D$, the integrated vertex operator is easily computed to be
\eqn\intv{GV = \p\T^\a A_\a  +\Pi_z^m A_m  + D\Phi_\a W^\a +\Phi_\a D\T^\b \N_\b W^\a 
+  \Phi_\a\Pi^m_\k \p_m W^\a}
$$+ D\T^\a D\T^\b (-\half A_m \g^m_{\a\b} + \N_\b A_\a) +D\T^\a \Pi_\k^m (\p_m A_\a - \N_\a A_m)
 +  \Pi^m_\k \Pi^n_\k\p_m A_n.$$
The constraints of \consone\ imply that the $\k=0$ component of the superfields $\Phi_\a$ and $\Pi_\k^m$ carry $-1$ U(1) charge, and the condition that $(GV)_2=0$ implies
the d=10 super-Yang-Mills equation of motion $\g_{m_1 ... m_5}^{\a\b} \N_\a A_\b=0$.
So
\eqn\gvz{(GV)_0 = \p\t^\a A_\a +  \pi^m A_m + h_\a W^\a   + \l^\b\O_\a  \N_\b W^\a}
$$ +\l^\a \pi_n {{(\g^n \lb)^\b}\over{\l\lb}} (-A_m \g^m_{\a\b} +\N_\b A_\a+ \N_\a A_\b)  +\l^\a \Pi_\k^m (\p_m A_\a - \N_\a A_m)
.$$
Using the definitions of \deffO,
one can verify that $(GV)_0$ is independent of $\lb_\a$ if 
\eqn\indlb{\N_\a A_\b + \N_\b A_\a = \g^m_{\a\b} A_m, \quad  \N_\a A_m- \p_m A_\a  = 
\g_{m\a\b} W^\b,}
which are the usual onshell superfield constraints for d=10 super-Yang-Mills. And after imposing
these super-Yang-Mills constraints, 
$(GV)_0$ reproduces the pure spinor integrated vertex operator
\eqn\psvo{U=(GV)_0 =\p\t^\a A_\a  +  \pi^m A_m + d_\a W^\a  +{1\over 4}(w\g^{mn} \l) F_{mn} }
where $D_\a W^\b ={1\over 4} (\g^{mn})^\b{}_\a F_{mn}$ and $d_\a \equiv
h_\a -(\l\g_m)_\a \Pi_\k^m$ differs from the definition of $d_\a$ in \deffO\
by a term with $-2$ U(1) charge which does not contribute to \psvo. Note that $(GV)_{-2}$ is nonzero and
satisfies $G^- \int (GV)_0 =-  G^+\int (GV)_{-2}$. This explains why the usual pure spinor integrated
vertex operator $U$ of \psvo\ is not annihilated by the $b$ ghost \ref\BakhmatovFPA{
  I.~Bakhmatov and N.~Berkovits,
  ``Pure Spinor $b$-ghost in a Super-Maxwell Background,''
JHEP {\bf 1311}, 214 (2013).
[arXiv:1310.3379 [hep-th]].
} but satisfies $b_{-1} \int U= Q \int\Lambda$
where $\Lambda =- (GV)_{-2}$.\foot{An interesting question is which vertex operators satisfy
$(GV)_{-2}=0$ and therefore are annihilated by the $b$ ghost and preserve N=2 worldsheet supersymmetry. By analyzing \intv, one finds that $(GV)_{-2}=0$ if and only if $\lb_\a W^\a=0$.
Since $D_\b W^\a = {1\over 4} (\g^{mn})^\a{}_\b F_{mn}$, $\lb_\a W^\a=0$ implies that 
$F_{mn} (\g^{mn}\lb)_\a=0$, so $\lb_\a$ is a killing spinor in these backgrounds. I would like to thank Andrei Mikhailov for discussions on this point.}

\subsec{Tree-level scattering amplitudes}

In the pure spinor formalism, the usual tree-level $N$-point open string scattering amplitude prescription is to take 3 vertex operators $V_r$ of ghost-number one and conformal weight zero, and $N-3$
integrated vertex operator $U_r$ of ghost-number zero and conformal weight one. 
One then defines the tree amplitude $A$ to be the correlation function 
\eqn\treepure{A= \langle V_1 (z_1) V_2 (z_2) V_3 (z_3) \int dz_4 U_4 ... \int dz_N U_N \rangle}
where the $(z_1, z_2, z_3)$ are arbitrary points and the zero mode normalization is defined by 
$\langle (\l\g^m\t)(\l\g^n\t)(\l\g^p\t)(\t\g_{mnp}\t)\rangle =1$. Although this prescription only
requires 5 of the 16 $\t$ zero modes to be present in the integrand, it is spacetime supersymmetric since one can show that any term in the integrand with more than 5 $\t$ zero modes and $+3$ ghost-number is not in the cohomology of $Q=\int \l^\a d_\a$ \BerkovitsFE. 

But before twisting, the vertex operators $V_r$ of $+1$ ghost-number have conformal weight $+\half$. So this prescription is only conformally invariant after twisting the pure spinor formalism and is inconsistent in the untwisted pure spinor formalism.
Fortunately, there is an alternative prescription one can define for tree-level amplitudes in the pure spinor formalism which only involves ghost-number zero vertex operators $U$ and can be defined both before and
after twisting.

In this alternative prescription, one takes $N$ integrated vertex operators $U_r$ of ghost-number zero and conformal weight one and defines the tree amplitude as
\eqn\treepuretwo{A= \langle (z_1 - z_2)(z_2-z_3)(z_3-z_1) U_1 (z_1) U_2 (z_2) U_3 (z_3) \int dz_4 U_4 ... \int dz_N U_N \rangle}
where $(z_1, z_2, z_3)$ are arbitrary points and the zero mode normalization is defined by 
$\langle 1 \rangle =1$. In this prescription, none of the 16 $\t$ zero modes need to be present in the integrand. But it is again spacetime supersymmetric since one can show that the only term in the cohomology of $Q=\int \l^\a d_\a$ with zero ghost-number is the identity operator.
So any term with $\t$ zero modes and zero ghost number will decouple since it is not in the
cohomology of $Q= \int \l^\a d_\a$. 

For example, consider the $N$-point Yang-Mills tree amplitude where $U$ is defined in \psvo.
For $N$ external gluons, requiring an equal number of $\t^\a$ and $p_\a$ zero modes
implies that the only term in \psvo\ which contributes is
\eqn\checku{U = \p x^m A_m (x) + \half  M^{mn} F_{mn}(x)}
where $M^{mn} =\half (p\g^{mn}\t) + \half (w\g^{mn}\l)$. Using the fact that $M^{mn}$
is a Lorentz current of level 1 with the same OPE's as the RNS Lorentz current
$\psi^m \psi^n$, one can easily verify that the prescription of \treepuretwo\
reproduces the correct tree amplitudes.

A similar zero mode prescription was used by Lee and Siegel in \ref\LeePA{
  K.~Lee and W.~Siegel,
  ``Simpler superstring scattering,''
JHEP {\bf 0606}, 046 (2006).
[hep-th/0603218].
}, and is closely related to
the ${\cal F}_1$ picture for scattering Neveu-Schwarz states in the RNS formalism. To compute $N$-point open string RNS tree amplitudes in this ${\cal F}_1$ picture, one chooses all $N$ Neveu-Schwarz vertex operators in the zero picture and uses the same zero mode regularization
$\langle c(z_1) c(z_2) c(z_3) \rangle =  (z_1 - z_2)(z_2-z_3)(z_3-z_1)$ as in the bosonic string.
Although it is unclear how to generalize this prescription to loop amplitudes in the RNS formalism, it is easy to show that computations in the ${\cal F}_1$ picture reproduce the same tree-level amplitude prescription as in
the conventional ${\cal F}_2$ picture where two Neveu-Schwarz vertex operators
are chosen in the $-1$ picture and one uses the zero mode regularization
\eqn\fone{\langle c(z_1) e^{-\phi(z_1)} c(z_2) e^{-\phi(z_2)} c(z_3)   \rangle =  (z_1 - z_3)(z_2-z_3).}
To prove the equivalence of RNS computations in the ${\cal F}_1$ and ${\cal F}_2$
pictures, multiply the BRST-invariant state $(c \p c \p^2 c e^{-2\phi})$ appearing in \fone\ with two picture raising operators
to obtain a BRST-invariant state of ghost-number three and
zero picture which includes the term $c\p c \p^2 c$. 

So for computing tree-level scattering amplitudes in the untwisted pure spinor formalism,
the prescription of \treepuretwo\ can be used. Because of U(1) charge conservation with respect to $J = 
-\l^\a w_\a$ and the absence of terms with positive U(1) charge in the worldsheet action and
vertex operators, terms with negative U(1) charge
cannot contribute to tree amplitudes using this precription.
Furthermore, it will be verified in the next section that the tree
amplitude prescription of \treepuretwo\ for Neveu-Schwarz states in the extended RNS formalism gives the same tree amplitudes as in the usual RNS formalism.
Although it will not be verified here, it is natural to conjecture that the prescription of \treepuretwo\ with the zero mode normalization $\langle 1\rangle =1$ can also be used to compute twistor-string
tree amplitudes.

\subsec{Extended RNS formalism}

The N=1 worldsheet superfields in the pure spinor formalism are
$[X^m, \T^\a, \Phi_\a]$ satisfying the constraints of \consone\ that
$(\g^m \lb)^\a (D X_m -\half D \T \g_m \T) =0$ and $(\lb \g^m\Phi)  =0$.
If one shifts $\T^\a$ by defining
\eqn\defK{\T'^\a \equiv \T^\a + K_m (\g^m\lb)^\a ~~{\rm where}~~
K^m = - {1\over{\lb D\T'}}DX^m + \half  {{\lb\T'}\over{\lb D\T'}} D ( {{DX^m}\over{\lb D\T'}}),}
$(\g^m \lb)^\a (D X_m - \half D\T \g_m \T) =0$ implies that 
$(\lb\g_m)^\a (\T'\g^m D\T') =0.$
So the N=1 superfields $[\T'^\a, \Phi_\a]$ satisfy the constraints
\eqn\consrns{(\lb\g_m)^\a(\T'\g^m D\T') =0, \quad \lb\g^m\Phi =0,} 
and leave unconstrained the N=1 superfield $X^m$. Interpreting $X^m$ as the usual N=1
worldsheet superfield of the RNS formalism, the formalism including both
$X^m$ and $(\T'^\a, \Phi_\a)$ will be called the ``extended RNS formalism''. In components, it is convenient to expand the new superfield $\T'^\a$ as
\eqn\expansionTh{
\T'^\a = \t'^\a + \k\lambda^\a + (\g^m\lb)^\a ( f_m +\k g_m)}
where $\t'^\a$ and $\l^\a$ are constrained to satisfy
\eqn\constrth{\t'\g^m \l =0, \quad \l\g^m\l =0,}
and $(\lb\g_m)^\a (\T'\g^m D\T') =0$ implies that $f_m$ and $g_m$ are quadratic and higher-order in $\t'^\a$.

In terms of the N=(1,0) superfields $[X^m, \T'^\a, \Phi_\a]$, the heterotic worldsheet action in the extended RNS formalism will be defined to be the sum of the usual RNS action with
an action for the $[\T'^\a, \Phi_\a]$ superfields 
as
\eqn\rnshet{S = \int d^2 z d\k [ \half DX^m \bar\p X_m +\Phi_\a \bar \p\T'^\a 
+ (\lb\g_m L)(\T'\g^m D\T') +M_m (\lb\g^m\Phi)] .} And in terms of the N=(1,1) superfields
$[X^m, \T'^\a, \Th'^\ah, \Phi_\a, \widehat \Phi_\ah]$, the Type II worldsheet action will be defined as
\eqn\rnstwo{S = \int d^2 z d\k d\kb [ \half DX^m \bar DX_m- \Phi_\a \bar D\T'^\a + \widehat\Phi_\ah D\Th'^\ah }
$$+ (\lb\g_m L)(\T'\g^m D\T') +M_m (\lb\g^m\Phi) 
+ (\lbh\g_m \widehat L)(\Th'\g^m \bar D\Th')
+\widehat M_m (\lbh\g^m\widehat\Phi)
].$$

To relate the pure spinor heterotic action of \het\ to the extended RNS heterotic action of \rnshet, substitute into \hetlast\ the component form of \defK\ which is
\eqn\comK{\t^\a = \t'^\a - [{1\over{\lb\l}}\psi^m - {(\lb\t')\over{2(\lb\l)^2}} (\p x^m - \psi^m {{(\lb\p\t')}\over{(\lb\l)}}) +f^m](\g_m\lb)^\a, }
where $f_m$ is quadratic and higher-order in $\t'^\a$.
In terms of $\t'^\a$, the action of \hetlast\ is 
\eqn\hetlasttwo{S = \int d^2 z [  h_\a \bar\p\t'^\a + \O_\a\bar\p\l^\a  - \psi_m (\l\g^m\g^n\lb)
\bar\p ({1\over{\l\lb}} \psi^n - {(\lb\t')\over{2(\lb\l)^2}} \p x^n  ) +\half\p x^m \bar\p x_m +  O(\t')   ]}
$$= \int d^2 z [  (h_\a -\lb_\a r^m \p x_m) \bar\p\t'^\a + (\O_\a -2 \psi_m r^m \lb_\a) \bar\p\l^\a  - \psi_m \bar\p\psi^m +\half\p x^m \bar\p x_m +  O(\t')  ]$$
where $O(\t')$ denotes terms linear or higher-order in $\t'^\a$ (counting its conjugate momentum $h_\a$ 
as an inverse power of $\t'^\a$) and
\eqn\defrm{r^m = \psi_ n {{\lb \g^m\g^n\l}\over{2(\l\lb)^2}}.}
Defining $\Phi_\a = \tilde\O_\a + \k \tilde h_\a$ in the extended RNS action of \rnshet\ where
\eqn\defoprime{\tilde\O_\a\equiv \O_\a -2 \psi_m r^m \lb_\a,
\quad \tilde h_\a \equiv h_\a - r^m \p x_m \lb_\a,}
one can easily verify that \rnshet\ reproduces \hetlasttwo\ if one ignores terms proportional to 
$O(\t')$.

To understand why these $O(\t')$ terms can be ignored, note that physical vertex operators
will be required to carry zero U(1) charge with respect to $J=- \l^\a \tilde\O_\a$
and be globally defined on the pure spinor
space, i.e. physical vertex operators must be independent of $\lb_\a$ and be invariant under the gauge transformations
\eqn\gaugerns{ \d\tilde \O_\a = \xi_m (\g^m \l)_\a + \rho_m (\g^m \t')_\a, \quad  \d \tilde h_\a = 
(\g^m \l)_\a \rho_m,}
generated by the constraints of \constrth\ where $\xi_m$ and $\rho_m$ are arbitrary
parameters.
It can be verified that all quantities
with zero U(1) charge which are gauge-invariant under \gaugerns\ must contain non-negative powers of $\t'^\a$ (where $\tilde h_\a$ counts as an inverse power of $\t'^\a$), e.g.
the Lorentz current $M^{mn} = \half (\tilde\O\g^{mn}\l +\tilde h\g^{mn}\t')$. And since the tree amplitude prescription
vanishes unless there are an equal number of $\t'^\a$'s and $\tilde h_\a$'s in the correlation
function, one can ignore any $O(\t')$ terms in the worldsheet action which are linear or higher-order
in $\t'^\a$.

Finally, it will be argued that the computation of tree-level scattering amplitudes of physical states 
using the prescription of \treepuretwo\ in the extended RNS formalism is equivalent to the computation of Neveu-Schwarz states using the usual RNS prescription in the ${\cal F}_1$ picture. To prove this equivalence, one needs to show that the extra fields $(\l^\a, \t'^\a, \tilde\O_\a, \tilde h_\a)$ in
the extended RNS formalism do not contribute to tree-level scattering amplitudes using the
zero mode normalization where $\langle 1 \rangle =1$. 

Any physical vertex operator in the extended RNS formalism must be gauge-invariant under \gaugerns\ and be a worldsheet primary field with zero U(1) charge. Examples of such
gauge-invariant operators with zero U(1) charge which involve $\tilde h_\a$ and $\tilde\O_\a$ are
$M^{mn} =\half (\tilde\O\g^{mn}\l +\tilde h\g^{mn}\t')$ and its
derivatives.
But since $M^{mn}$ has level zero, i.e. the OPE of $M^{mn}$ with $M^{pq}$ has no double
pole proportional to the identity operator, it is not possible for a correlation function involving $M^{mn}$ and its derivatives to be proportional
to the identity operator. It seems
reasonable to conjecture that all gauge-invariant opeartors depending on $\tilde h_\a$ or $\tilde\O_\a$ are of this type and cannot produce the identity operator in their OPE's. Therefore, any terms in the vertex operator which depend on $\tilde h_\a$ or $\tilde\O_\a$ 
will decouple from the tree amplitudes. Furthermore, since the tree amplitude vanishes unless there are an equal number of $\tilde h_\a$'s and $\t'^\a$'s in the correlation function, 
any terms in the vertex operator which depend on $\t'^\a$ will also decouple. So the only
terms in the vertex operator which can contribute to tree amplitudes are terms that only depend
on the superfield $X^m$. But worldsheet N=1 superconformal primary fields which only
depend on $X^m$ are the usual Neveu-Schwarz states in the RNS formalism. 
So tree amplitudes of physical states in the extended RNS formalism are equivalent to the
tree amplitudes of Neveu-Schwarz states in the usual RNS formalism. 

\subsec{Worldsheet supersymmetric action in curved background}

By adding integrated vertex operators to the worldsheet action in a flat target-space background,
one can generalize the N=1 worldsheet supersymmetric actions to a
curved background. For the
heterotic superstring in the pure spinor description, the N=(1,0) worldsheet supersymmetric action of \het\ generalizes in an N=1 d=10 supergravity background to
\eqn\hetc{S = \int d^2 z d\k [\half \eta_{ab} E_M^a DZ^M E_N^b \bar\p Z^N + \half
B^{het}_{MN} DZ^M \bar\p Z^N + \Phi_\a E^\a_M \bar\p Z^M }
$$+ (\lb\g_a L)E_M^a DZ^M + M_a (\lb\g^a \Phi) ]$$
where $M = (m,\mu)$ are curved-space indices for $m=0$ to 9 and $\mu=1$ to 16,
$A = (a,\a)$ are tangent-space indices for $a=0$ to 9 and $\a=1$ to 16, $Z^M = (X^m,\T^\mu)$, $E_M^A$ is the super-vierbein, and $B^{het}_{MN}$ is the graded antisymmetric tensor superfield.

And for the Type II superstring in the pure spinor description, the N=(1,1) worldsheet
supersymmetric action of \twoact\ generalizes in an N=2 d=10 supergravity background
to 
\eqn\closedc{S = \int d^2 z d\k d\bar\k [\half \eta_{ab} E_M^a DZ^M E_N^b \bar D Z^N + 
\half B^{II}_{MN} DZ^M \bar D Z^N }
$$- \Phi_\a E^\a_M \bar D Z^M + \widehat\Phi_\ah E^\ah_M  D Z^M -
F^{\a\ah}\Phi_\a \widehat\Phi_\ah  $$
$$+ (\lb\g_a L) E^a_M DZ^M +M_a (\lb\g^a\Phi) 
+ (\lbh\g_a \widehat L) E^a_M \bar D Z^M
+\widehat M_a (\lbh\g^a\widehat\Phi) ],$$
where  $M = (m,\mu, \hat\mu)$ are curved-space indices,
$A = (a,\a, \ah)$ are tangent-space indices, $Z^M = (X^m,\T^\mu, \widehat\T^{\hat\mu})$, $E_M^A$ is the super-vierbein, $B^{II}_{MN}$ is the graded antisymmetric tensor superfield, and $F^{\a\ah}$
is the superfield whose lowest components are the
Type II Ramond-Ramond bispinor field strengths.

After imposing the constraints from varying the Lagrange multipliers, one can expand the actions of \hetc\ and
\closedc\ in components in terms of the pure spinor worldsheet variables $(Z^M, d_\a, \l^\a, w_\a, \lb_\a)$. Although it will not be verified here, it is expected that when the supergravity fields
are onshell, all terms in the action will have either zero or negative U(1) charge with respect
to \jgen, and the terms with zero U(1) charge will be independent of $\lb_\a$ and reproduce the pure spinor worldsheet
action in a curved background of \ref\howe{N.~Berkovits and P.~S.~Howe,
  ``Ten-dimensional supergravity constraints from the pure spinor formalism for the superstring,''
Nucl.\ Phys.\ B {\bf 635}, 75 (2002).
[hep-th/0112160].}. 

Using the extended RNS description, the heterotic superstring action of \rnshet\ can be
generalized in a Neveu-Schwarz background and the resulting action is
\eqn\rnshetb{S = \int d^2 z d\k [ \half (g_{mn}(X) + b_{mn}(X)) DX^m \bar\p X_m +  \Phi_\a (\bar \nabla_{\bar z} \T')^\a}
$$ 
+ (\lb\g_a L) (\T'\g^a \nabla_\k\T') +M_a (\lb\g^a\Phi)] $$
where $(\bar\nabla_{\bar z}\T')^\a = \bar\p\T'^\a +\bar\p X^m \omega_{m\b}{}^\a (X) \T'^\b$,
$(\nabla_\k\T')^\a = D\T'^\a +D X^m \omega_{m\b}{}^\a (X) \T'^\b$,
and
$\omega_{n\b}{}^\a$ is the spin connection. 
Similarly, the Type II worldsheet action of \rnstwo\ generalizes in a Neveu-Schwarz/Neveu-Schwarz
background to 
\eqn\rnstwob{S = \int d^2 z d\k d\kb [\half (g_{mn}(X)+b_{mn}(X))  DX^m \bar DX^n - \Phi_\a (\bar\nabla_{\bar\k}\T')^\a + \widehat\Phi_\ah (\nabla_\k\Th')^\ah }
$$+ (\lb\g_a L)(\T'\g^a \nabla_\k\T') +M_a (\lb\g^a\Phi) 
+ (\lbh\g_a \widehat L)(\Th'\g^a \bar \nabla_{\bar \k}\Th')
+\widehat M_a (\lbh\g^m\widehat\Phi)
],$$
where 
$(\bar\nabla_{\bar\k}\T')^\a = \bar D\T'^\a +\bar D X^m \omega_{m\b}{}^\a (X) \T'^\b$,
$(\nabla_\k\Th')^\ah = D\Th'^\ah +D X^m \widehat\omega_{m\bh}{}^\ah (X) \Th'^\bh$,
and
$\omega_{n\b}{}^\a$ and $\widehat\omega_{n\bh}{}^\ah$ are the
left and right-moving spin connections.

\newsec{Twistor String Formalism}

\subsec{Twistor superfields}

By choosing different solutions of the superfield constraints of \consone, 
\eqn\consa{(\g^m \lb)^\a (D X_m -\half D \T \g_m \T) =0, \quad (\lb \g^m)^\a \Phi_\a  =0,}
one obtains
different worldsheet supersymmetric descriptions of the superstring. Expanding the superfields in component fields as
\eqn\offst{X^m = x^m + \k \psi^m, \quad \T^\a = \t^\a + \k \L^\a, \quad \Phi_\a = \O_\a + \k h_\a,}
the pure spinor description solves for $\psi^m$ and $h_\a$ in terms of $d_\a$ through
the equations \superx\ and \deffO. And in the extended RNS description,
one solves for $\t^\a$ in terms of $\psi^m$ and a constrained $\t'^\a$ satisfying $(\l\g^m\t')=0$ by shifting $\T^\a$ to $\T'^\a$ as in \defK. In both of these descriptions, the bosonic component fields $\L^\a$ and $\O_\a$ are solved in terms of $x^m$ and $(\l^\a, w_\a)$ where $\l^\a$ is a pure spinor and
\eqn\supersol{\L^\a = \l^\a + {1\over{2(\l\lb)}} (\p x^m - \half\p\t\g^m \t) (\g_m\lb)^\a, \quad 
\O_\a = w_\a - {1\over{2(\l\lb)}} (\lb\g^m w)(\g_m\l)_\a. } 

In this section, a new twistor-like solution for the constraints of \consa\ will be presented in which the superfield $X^m = x^m + \k\psi^m$ is solved in terms of the other superfields. 
In this description, the superfield $\Phi_\a$ is shifted to 
$\Phi'_\a = \Phi_\a - \half X^m (\g_m D\T)_\a$ 
whose components $\Phi'_\a \equiv \O'_\a + \k h'_\a$ no longer satisfy $\lb\g^m\O' = \lb\g^m h'=0$. It will be convenient to expand
the bosonic component fields $\L^\a$ and $\O'_\a$ which appear in $\T^\a$ and $\Phi'_\a$
as 
\eqn\defmu{\L^\a = \l^\a + {1\over{2(\l\lb)}} (\g^m \lb)^\a (\l\g_m\nu), \quad
\O'_\a = \mu_\a + w_\a -   {1\over{2(\l\lb)}}  (\g^m\l)_\a  (w \g_m\lb ),}
where $\l^\a$ and $\mu_\a$ are constrained to satisfy
\eqn\mucon{\l\g^{mn}\mu = \mu\g^m\mu = \l\g^m\l=0.}

The variables  $w_\a$ and $\nu^\a$ in \defmu\ are the conjugate momenta to $\l^\a$ and $\mu_\a$
and are defined up to the gauge transformations
\eqn\gaugetrmu{
\d w_\a =  f_m (\g^m \l)_\a, \quad
\d \nu^\a = c_{mn} (\g^{mn}\l)^\a + h_m (\g^m \mu)^\a, }
for arbitrary parameters $c_{mn}$, $h_m$ and $f_m$.
Note that $(\l^\a, \mu_\a)$ satisfying \mucon\ contain 16 independent
components and can be interpreted as d=10 twistor variables. As discussed in \ref\BerkovitsBW{
  L.P. Hughston, ÒThe Wave Equation in Even Dimensions,Ó in Further Advances in
Twistor Theory, vol. 1, Research Notes in Mathematics 231, Longman, pp. 26-27,
1990\semi
  N.~Berkovits and S.~A.~Cherkis,
  ``Higher-dimensional twistor transforms using pure spinors,''
JHEP {\bf 0412}, 049 (2004).
[hep-th/0409243].},
twistors in $d$ spacetime dimensions are pure spinors in $d+2$ dimensions which transform covariantly under $SO(d,2)$ conformal transformations. The spinors $(\l^\a, \mu_\a)$ satisfying \mucon\ can therefore
be interpreted as the 16 independent components of a d=12 pure spinor $U^A$ satisfying
the pure spinor condition $U^A \g^{MN}_{AB} U^B$ where $A=1$ to 32, $M=0$ to 11, and $\g^M$
are the $d=12$ gamma-matrices. So \defmu\ decomposes
the 32 components of $\L^\a$ and $\O'_\a$ into the 16 independent components of $(\l^\a, \mu_\a)$ describing a d=12 pure spinor, and the 16 gauge-invariant components of
its conjugate momenta $(w_\a, \nu^\a)$.


For the heterotic superstring, $\Phi'_\a$ does not have enough degrees of freedom
to solve for $X^m$, but for the Type II superstring, one also has the right-moving superfields
 \eqn\rhtw{\Th^\ah = \th^\ah + \bar\k \Lh^\ah, \quad \widehat\Phi'_\ah = \widehat \O'_\ah + \bar\k \widehat h'_\ah,}
 with component expansions
 \eqn\defmuhat{\widehat\L^\ah = \lh^\a + {1\over{2(\lh\lbh)}} (\g^m \lbh)^\ah (\lh\g_m\widehat\nu), \quad
\widehat\O'_\ah = \widehat\mu_\ah + \widehat w_\ah -   {1\over{2(\lh\lbh)}}  (\g^m\lh)_\ah  (\widehat w \g_m\lbh ).}
The shifted superfields $\Phi'_\a$ and $\widehat\Phi'_\ah$ are defined by
\eqn\phip{\Phi'_\a = \Phi_\a -\half X^m (\g_m D\T)_\a, \quad \widehat\Phi'_\ah = \widehat\Phi_\ah - \half X^m (\g_m \bar D\Th)_\ah, } 
and no longer satisfy the constraints
$\lb\g^m\Phi'=0$ and $\lbh\g^m\widehat\Phi'=0$. Since the $\Phi'_\a$ and $\widehat\Phi'_\ah$ superfields are related to the $X^m$ superfield by 
\eqn\relphit{\lb\g^m\Phi' =-\half (\lb \g^m \g^n D\T) X_n, \quad \lbh\g^m\widehat\Phi' =-\half (\lbh \g^m \g^n \bar D\Th) X_n,}
the bosonic spinor variables $(\mu_\a, \l^\a)$ and $(\widehat\mu_\ah, \lh^\ah)$ in \defmu\ and \defmuhat\ are related
to the spacetime vector variable $x^m$ by the usual twistor relation
\eqn\twis{\mu_\a = - \half x^m (\g_m\l)_\a, \quad \widehat\mu_\ah = - \half  x^m (\g_m\lh)_\ah.}

For the Type IIA superstring, the twistor relation of \twis\ can be inverted to solve for
$x^m$ in terms of $\mu_\a$ and $\widehat\mu^\a$ as
\eqn\solvex{x^m = - {1\over{(\l\lh)}}(\lh \g^m \mu + \l\g^m \widehat\mu)}
where it is assumed that $\l^\a \lh_\a \neq 0$. Similarly, the Type IIA superfield $X^m$ can be
expressed in terms of $\Phi'_\a$ and $\widehat\Phi'^\a$ as
\eqn\simsuper{X^m =- {1\over{(D\T \bar D\Th)}}(\bar D\Th \g^m \Phi' + D\T\g^m \widehat\Phi').}
Although there is no analogous
solution for the uncompactified Type IIB superstring, one can use
the standard T-duality relation of Type IIB with Type IIA to solve for $x^m$ if at least one direction of the Type IIB superstring is compactified on a circle. For example, if $x^9$ is compactified on a circle, define
\eqn\tdualmu{\mu_\a =-\half \xwt^m (\g_m \l)_\a, \quad \widehat\mu_\a = - \half  \xwt^m (\g^9\g_m \g_9 \lh)_\a,}
where $\xwt^m$ is the T-dual to $x^m$ defined by 
\eqn\defxwt{\xwt^m = x_L^m + x_R^m ~~ {\rm for}~~ m=0 {\rm~~ to~~} 8, \quad
\xwt^9 = x_L^9 - x_R^9,}
and $x^m_L$ and $x^m_R$ are the left and right-moving
parts of $x^m$ defined by $x^m_L(z) = \int^z dy \p x^m(y)$ and
$x^m_R(\bar z) = \int^{\bar z} d\bar y \bar\p x^m(\bar y)$. Using \tdualmu, one can invert to solve for $\xwt^m$ in terms of
$\mu_\a$ and $\widehat\mu_\a$ as 
\eqn\tdualx{\xwt^m = - {1\over{(\l\g^9\lh)}} (\lh\g^9 \g^m \mu + \l\g^m \g^9 \widehat\mu)}
where it is assumed that $(\l\g^9\lh)\neq 0$.

\subsec{N=1 superconformal and U(1) generator} 

The left-moving N=1 superconformal stress tensor for the twistor-string is ${\cal T} = \half D\Phi'_\a D\T^\a -\half \Phi'_\a \p\T^\a$, which in 
components is
\eqn\stresstw{G =h'_\a \L^\a - \O'_\a \p\t^\a , \quad T =-\half \O'_\a\p\L^\a +\half \p\O'_\a \L^\a - h'_\a \p\t^\a.}
As in the other worldsheet supersymmetric descriptions of the superstring, physical states will
be required to be N=1 superconformal primary fields whose integrated
vertex operators have zero or negative
charge with respect to a U(1) generator $J$. In the twistor-string description, 
the U(1) generator will be
defined as 
\eqn\Jt{J = -\l^\a w_\a + \mu_\a \nu^\a }
which splits $G$ into 
\eqn\splitG{G^+ = h'_\a \l^\a - \mu_\a \p\t^\a, \quad G^- = {1\over{2(\l\lb)}}
(h'\g_m\lb)(\l\g^m\nu) -  w_\a \p\t^\a + {1\over{2(\l\lb)}} (w\g_m\lb)(\l\g^m\p\t)}
where $(\l^\a, \mu_\a, w_\a, \nu^\a)$ are defined in \defmu.
Note that the U(1) generator $J$ of \Jt\ counts the projective weight of the d=10 twistor variables
where $(\l^\a, \mu_\a)$ carry projective weight $+1$ and $(w_\a, \nu^\a)$ carry
projective weight $-1$. So the integrated vertex operator $GV$ will be required to carry zero or negative projective weight, and the term $(GV)_0$ of zero projective weight will
be required to be globally defined on pure spinor space, i.e. independent of $\lb_\a$
and invariant under the gauge transformations of \gaugetrmu.

\subsec{Worldsheet action in a flat background}

Under spacetime supersymmetry, \relphit\ and $\d X^m = -\half(\e \g^m\T + \widehat\e\g^m\Th)$ implies that the $\Phi'_\a$ and $\widehat\Phi'_\ah$
superfields transform as 
\eqn\susytw{\d\T^\a = \e^\a, \quad \d \Th^\ah = \widehat\e^\ah,}
$$
\d\Phi'_\a ={1\over 4} (\e\g^m\T + \widehat\e\g^m\Th) (\g_m D\T), \quad
\d\widehat\Phi'_\ah = {1\over 4}(\e \g^m\T + \widehat\e\g^m\Th) (\g_m\bar D\Th).$$
And under spacetime translations, $\d X^m = c^m$ implies that the $\Phi'_\a$ and $\widehat\Phi'_\ah$
superfields transform as 
\eqn\transtw{\d\T^\a = \d \Th^\ah =0,\quad
\d\Phi'_\a =-\half c^m (\g_m D\T), \quad
\d\widehat\Phi'_\ah = -\half c^m (\g_m\bar D\Th).}
The N=(1,1) worldsheet supersymmetric action for the Type II twistor-string in a flat background should be invariant under these super-Poincar\'e transformations and will be defined in terms of the $(\T^\a,\Th^\ah, \Phi'_\a, \widehat\Phi'_\ah)$ superfields
as
\eqn\twtwo{S = \int d^2 z d^2 \k [-\Phi'_\a \bar D \T + \widehat\Phi'_\ah D \Th
- {1\over 8}(\T\g^m D\T)(\Th\g_m \bar D\Th) +{1\over 8} (\Th\g^m D\Th)(\T\g_m\bar D\T)].}

After integrating out auxiliary variables and shifting $w_\a$ and $\widehat w_\ah$, \twtwo\ reduces to
 \eqn\twcom{S = \int d^2 z [h'_\a \bar \p \t^\a + \widehat h'_\ah \p \th^\ah +
 w_\a \bar\p \l^\a + \widehat w_\ah \p \lh^\ah + \mu_\a \bar\p \nu^\a + \widehat \mu_\ah \p \widehat \nu^\ah }
 $$-\half (\nu\g^m \l - \half\t \g^m \p\t)(\widehat \nu \g_m \lh  -\half \th \g_m \bar\p \th)],$$
 with the spacetime supersymmetry generators 
 \eqn\susycomp{q_\a =\int dz d\k \Phi'_\a +{1\over 4}\int d\bar z d\bar\k (\g^m\T)_\a(\Th\g_m\bar D\Th)}
 $$=
  \int dz h'_\a
 -\half \int d\bar z  (\lh \g^m \widehat \nu -\half \th \g^m\bar\p\th)(\g_m\t)_\a ,$$
 $$\widehat q_\ah ={1\over 4} \int dz d\k (\g^m\Th)_\ah (\T\g_m D\T) + \int d\bar z d\bar \k \widehat\Phi'_\ah$$
 $$=-\half\int d z  (\l \g^m  \nu - \half\t \g^m\p\t)(\g_m\th)_\ah +
  \int d \bar z \widehat h'_\ah ,$$
  $$P_m =\half \int dz d\k \T\g_m D\T +\half \int d\bar z d\bar\k \Th \g_m \bar D\Th$$
  $$ = \int dz (\l\g_m \nu -\half\t\g_m \p\t) + \int d \bar z (\lh\g_m \widehat \nu -\half \th \g_m \bar\p\th).$$

Using that
$\Phi'_\a + \half X^m (\g_m D\T)_\a$ and $\widehat\Phi'_\ah +\half X^m (\g_m \bar D\Th)_\ah$
are spacetime supersymmetric, the action of \twtwo\ can be written in manifestly spacetime
supersymmetric notation as
\eqn\twthree{ S = \int d^2 z d^2 \k [- (\Phi'_\a + \half X^m (\g_m D\T)_\a) \bar D \T^\a
 + (\widehat\Phi'_\ah  +\half X^m (\g_m \bar D\Th)_\ah) D \Th^\ah + B^{II}_{\k\bar\k}]}
 where $B^{II}_{\k\kb}$ is defined in \btwo.
Note that this action is related to the Type II worldsheet action of \twoact\ by dropping the term
\eqn\drop{\half\int d^2 z d\k d\bar\k ~\Pi_\k^m \bar\Pi_{\bar\k m}.}
Since $ \Pi_\k^m \bar\Pi_{\bar\k m}$ has left and right-moving U(1) charge $(-1,-1)$, the term of zero
U(1) charge in \drop\ can be expressed as the BRST-trivial term
\eqn\btriv{-\half \int d^2 z~ \oint G^+ \oint \widehat G^+ ( \Pi_\k^m \bar\Pi_{\bar \k m} ).}
So if all vertex operators are annihilated by $\oint G^+$ and $\oint\widehat G^+$, it seems reasonable to assume that dropping the term of \btriv\ will
not affect the scattering amplitudes since one can pull the countour integrals of
$G^+$ and $\widehat G^+$ off of the surface. However, since vertex operators have not yet been constructed in the twistor-string formalism, this assumption has not yet been verified by explicit computations.

To relate \twtwo\ with the usual component form of the Type IIA pure spinor worldsheet action,
substitute $\mu_\a =-\half x^m (\g_m\l)_\a$ and $\widehat\mu^\a = -\half x^m (\g_m\lh)^\a$ into \twcom\ and vary $\nu^\a$ and $\widehat\nu_\a$ to obtain the equations
of motion
\eqn\eomnu{(\lh\g^m\widehat\nu)(\g_m\l)_\a = \bar\pi^m (\g_m\l)_\a, \quad
(\l\g^m\nu) (\g_m\lh)^\a = \pi^m (\g_m\lh)^\a, }
which implies
\eqn\aux{(\widehat\l\g^m\widehat\nu) = {{(\widehat\l\g^m\g^n\l)}\over {2\l\lh}} \bar\pi_n, \quad
(\l\g^m\nu) ={{(\l\g^m\g^n\widehat\l)}\over{2\l\lh}} \pi_n}
where the equations of motion $\bar\p\l^\a = \p\lh_\a=0$ have been used.
Plugging the auxiliary equations of \aux\ back into the action of \twcom\ and ignoring terms which vanish
when $\bar\p\l^\a = \p\lh_\a=0$ (and can therefore be cancelled by
an appropriate shift of $w_\a$ and $\widehat w^\a$), one finds that
\eqn\news{S =
 \int d^2 z [h'_\a \bar \p \t^\a + \widehat h'_\ah \p \th^\ah +
 w_\a \bar\p \l^\a + \widehat w_\ah \p \lh^\ah}
 $$
  +\half \pi^m \bar\pi_m - 
 {{(\l\g^m\g^n\lh)}\over{2\l\lh}}\pi_m\bar\pi_n
  -{1\over 8}  (\t\g_m\p\t )(\widehat\t\g^m\bar\p\widehat\t)
]$$
\eqn\fine{ = \int d^2 z [p_\a \bar \p \t^\a + \widehat p_\ah \p \th^\ah +
 w_\a \bar\p \l^\a + \widehat w_\ah \p \lh^\ah +\half\p x^m \bar\p x_m]}
 $$
-  \int d^2 z {{(\l\g^m\g^n\lh)}\over{2\l\lh}}\pi_m\bar\pi_n,$$
where 
\eqn\shiftp{p_\a = h'_\a + \half x_m (\g^m\p\t)_\a +{1\over 2}( \p x_m+{1\over 4}\t\g_m\p\t
+{1\over 8} \th\g_m\p\th) (\g^m\t)_\a, }
$$ \widehat p^\a = \widehat h'^\a + \half x_m
(\g^m\bar\p\widehat\t)^\a +{1\over 2} (\bar\p x_m +{1\over 4}\th\g_m\bar\p\th
+{1\over 8} \t\g_m\bar\p\t) (\g^m\widehat\t)^\a,$$
and the first line of \fine\ is the Type IIA worldsheet action of \purefreetwo.
Furthermore, \shiftp\ implies that
$G^+$ and $\widehat G^+$ of \splitG\ are mapped into the pure spinor BRST currents 
\eqn\newbrst{G^+ = \l^\a h'_\a - \mu_\a \p\t^\a \to
G^+ = \l^\a d_\a,}
$$\widehat G^+ = \widehat\l_\a \widehat h'^\a - \widehat\mu^\a \bar\p\widehat\t_\a \to
\widehat G^+ = \widehat\l_\a \widehat d^\a,$$
where 
\eqn\defdtwo{d_\a = p_\a -\half ( \p x^m + {1\over 4}\t\g^m\p\t) (\g_m\t)_\a, \quad
\widehat d^\a = \widehat p^\a -\half ( \bar\p x^m + {1\over 4}\widehat\t\g^m\bar\p\widehat\t) (\g_m\widehat\t)^\a}
are defined as in \susydef\ and the equations of motion $\bar\p\t^\a = \p\th_\a =0$ have been used.

Finally, note that 
\eqn\defstt{-\int d^2 z {{(\l\g^m\g^n\lh)}\over{2\l\lh}}\pi_m\bar\pi_n = 
\int d^2 z \oint G^+ \oint \widehat G^+ ({1\over{2\l\lh}} d_\a \widehat d^\a)}
in the second line of \fine\ is BRST-trivial. So up to this BRST-trivial term related to 
\btriv, the Type II twistor-string action of
\twtwo\ is equal to the pure spinor action of \purefreetwo. 

\subsec{Twistor string in $AdS_5\times S^5$ background}

In principle, one can obtain the worldsheet action for the twistor string in an $AdS_5\times S^5$ background by deforming the Type IIB action in a flat background with the
vertex operator for the Ramond-Ramond five-form field strength and computing the back-reaction. However, a simpler method is to find a $PSU(2,2|4)$-invariant action which
reduces in the large radius limit to the Type IIB twistor string action in a flat background of
\twtwo.

To describe the $AdS_5\times S^5$ background, it is convenient to start with the standard
representation of $AdS_5\times S^5$ where $g(Z)$ takes values in the
supercoset ${{PSU(2,2|4)}\over{SO(4,1)\times SO(5)}}$ and define the supervierbein
$E^A_M$ by
\eqn\defE{ g^{-1} \p g = E_M^A \p Z^M}
where $A=(a,\a,\ah)$ denote the 10 bosonic and 32 fermionic generators of  
${{PSU(2,2|4)}\over{SO(4,1)\times SO(5)}}$ and $Z^M = [x^m, \t^\mu, \th^{\hat\mu}]$. Using the notation $\eta_{\a\bh} = 
(\g^{01234})_{\a\bh}$ and $\eta^{\a\bh} = 
(\g_{01234})^{\a\bh}$, 
the Ramond-Ramond field strength is given by $F^{\a\bh} = \eta^{\a\bh}$ and one
can choose the gauge where 
\eqn\bfield{B_{MN}DZ^M \bar DZ^N = \eta_{\a\ah} (E_M^\a E_N^\ah + E_N^\a E_M^\ah)
DZ^M \bar DZ^N }
$$=  \eta_{\a\ah} ( (g^{-1}Dg)^\a (g^{-1}\bar Dg)^\ah
+(g^{-1} Dg)^\ah (g^{-1}\bar D g)^\a).$$

Just as the term $\half\int d^2 z \int d\k d\bar\k \Pi_\k^m \bar\Pi_{\bar\k m}$ of \drop\ 
was dropped from the twistor-string action in a flat background, the twistor-string action in an $AdS_5\times S^5$ background will be defined by dropping the analog of this term in the curved background action of \closedc 
\eqn\anal{\int d^2 z \int d\k d\bar\k\half\eta_{ab} E_M^a E_N^b DZ^M \bar DZ^N =\int d^2 z \int d\k d\bar\k \half\eta_{ab} (g^{-1}Dg)^a (g^{-1}\bar Dg)^b.}
So the twistor-string action is
\eqn\ads{S = r^2\int d^2 z d\k d\bar\k [ \half\eta_{\a\ah} ( (g^{-1}Dg)^\a (g^{-1}\bar Dg)^\ah
+(g^{-1} Dg)^\ah (g^{-1}\bar D g)^\a)}
$$-\Phi'_\a (g^{-1} Dg)^\a + \widehat\Phi'_\ah (g^{-1} \bar Dg)^\ah -
\eta^{\a\ah}\Phi'_\a \widehat\Phi'_\ah ]$$
where $r$ is the $AdS_5$ radius
and $X^m$ is determined in terms of $[\T^\a, \Th^\ah, \Phi'_\a, \widehat\Phi'_\ah]$ through
the $AdS_5\times S^5$ generalization of \simsuper. 

Since the $AdS_5\times S^5$ superstring is a Type IIB superstring, defining $X^m$ in terms of $[\T^\a, \Th^\ah, \Phi'_\a, \widehat\Phi'_\ah]$ will require T-dualizing one of the $AdS_5\times S^5$ directions as explained in \tdualmu. 
A convenient choice which will
hopefully be explored in a future paper is to T-dualize one of the $S^5$ directions, which
breaks the manifest $SU(4)$ R-symmetry to $SO(4)\times U(1)$. This is the same
manifest symmetry as the spin-chain, and T-dualizing in this direction means that
the spin-chain ground state $Tr(Z^n)$ where $Z$ is the scalar with $+1$ U(1) charge is described by a string with winding number $n$.


Integrating out the auxiliary superfields $\Phi'_\a$ and $\widehat\Phi'_\ah$ in \ads, the $AdS_5\times S^5$ twistor-string action simplifies to 
\eqn\adso{S =r^2 \int d^2 z d\k d\bar\k \half\eta_{\a\ah} [ (g^{-1}Dg)^\a (g^{-1}\bar Dg)^\ah
-(g^{-1} Dg)^\ah (g^{-1}\bar D g)^\a].}
Surprisingly, this action is not only invariant under the local transformation
$\d g = g\Omega$ when $\Omega\in SO(4,1)\times SO(5)$, it is also invariant under
$\d g = g\Omega$ when $\Omega\in SO(4,2)\times SO(6)$. So the bosonic elements of the coset can be gauged
away and the action of \adso\ can
be expressed as 
\eqn\adsp{S = r^2\int d^2 z d\k d\bar\k  \half\eta_{\a\ah} [ (G^{-1}DG)^\a (G^{-1}\bar DG)^\ah
-(G^{-1} DG)^\ah (G^{-1}\bar D G)^\a]}
where $G$ is a fermionic coset taking values in ${{PSU(2,2|4)}\over{SO(4,2)\times SO(6)}}$.\foot{This action has vanishing beta function since the coset is a symmetric space with $PSU(2,2|4)$ in the numerator and there is no WZW term.\ref\BerkovitsZQ{
  N.~Berkovits, M.~Bershadsky, T.~Hauer, S.~Zhukov and B.~Zwiebach,
  ``Superstring theory on AdS(2) x S**2 as a coset supermanifold,''
Nucl.\ Phys.\ B {\bf 567}, 61 (2000).
[hep-th/9907200].
}}

Using $SU(2,2)\times SU(4)$ notation, this action can be expressed as
\eqn\adsq{S =r^2 \int d^2 z d\k d\bar\k (G^{-1}DG)^J_R (G^{-1}\bar DG)^R_J}
where $J=1$ to 4 is a spinor representation of $SO(4,2)=SU(2,2)$ and $R=1$ to 4 is a spinor
representation of $SO(6)=SU(4)$. Note that
the Maurer-Cartan equations imply that 
\eqn\mc{\int d^2 z d\k d\bar\k (G^{-1}DG)^J_R (G^{-1}\bar DG)^R_J = -
\int d^2 z d\k d\bar\k (G^{-1}DG)^R_J (G^{-1}\bar DG)^J_R}
up to a surface term. Defining the coset representative of $G$ as
\eqn\repG{G = e^{\T_R^J T^R_J} e^{\tilde\T_J^R T^J_R}}
where $T^J_R$ and $T^R_J$ are the generators of the 32 fermionic isometries,
the left-invariant forms are
\eqn\leftinv{(G^{-1} \p G)^J_R = \p \T^J_R, \quad
(G^{-1} \p G)^R_J = \p\tilde\T^R_J + \tilde\T^R_K \p\T^K_S \tilde\T^S_J,}
and under the 32 global fermionic isometries generated by 
$\d G = (\e^J_R T^R_J + \tilde\e^R_J T^J_R) G$, the superfields of \repG\ transform as
\eqn\susytr{\d\T_R^J = \e_R^J + \T^J_S\tilde \e^S_K \T^K_R, \quad 
\d\tilde\T_J^R = \tilde\e^R_J - \tilde\e^R_K \T^K_S \tilde\T^S_J - \tilde\T^R_K \T^K_S \tilde\e^S_J.}
Plugging in the left-invariant forms of \leftinv\ into the action of \adsq, one obtains the
remarkably simple action
\eqn\adsr{S = r^2\int d^2 z d\k d\bar\k (D\T^J_R \bar D\tilde\T^R_J +
D\T^J_R \tilde\T^R_K \bar D\T^K_S \tilde\T^S_J).}

To show that the action of \adsr\ reduces in the large radius limit to the flat space action of 
\twtwo, rescale 
\eqn\resc{\T^J_R \to {1\over{\sqrt r}}\T^J_R, \quad \tilde\T^R_J \to {1\over{\sqrt r}}\tilde\T^R_J, }
and write \adsr\ as
\eqn\adsu{
S = \int d^2 z d\k d\bar\k [-\Phi'_\a \bar D\T^\a + \widehat\Phi'_\ah D\Th^\ah
 -
{1\over {r}}\eta^{\a\ah} \Phi'_\a \widehat \Phi'_\ah +D\T^J_R \tilde\T^R_K \bar D\T^K_S \tilde\T^S_J],}
where in terms of the $\T^\a$ and $\Th^\ah$ superfields of 
\ads\ written in d=10 spinor notation with $\a,\ah=1$ to 16, $\T^J_R$ only involves the linear combination $\T^\a +i \Th^\ah$ and $\tilde\T^R_J$ only involves the linear combination $\T^\a -i\Th^\ah$. After expressing the quartic term in \adsu\  in terms of $\T^\a$ and $\Th^\ah$,
\eqn\adsv{
S = \int d^2 z d\k d\bar\k [-\Phi'_\a D\T^\a + \widehat\Phi'_\ah \bar D\Th^\ah -
{1\over {r}} \eta^{\a\ah} \Phi'_\a \widehat \Phi'_\ah}
$$+e
(D\T \g^a \T + D\Th \g^a \Th) (\bar D\T \g_a \T + \bar D\Th \g_a \Th) $$
$$+
f_{abcdef}(D\T \g^{abc} \T + D\Th \g^{abc} \Th) (\bar D\T \g^{def}\T + \bar D\Th \g^{def} \Th)]$$
where $e$ and $f_{abcdef}$ are constants which are invariant under $SO(4,1)\times SO(5)$ transformations and come from expressing $D\T^J_R \tilde\T^R_K \bar D\T^K_S \tilde\T^S_J$ in d=10 notation.
When $r\to \infty$, the $ \eta^{\a\ah} \Phi'_\a \widehat \Phi'_\ah$ term drops out of \adsv\  and,
after appropriately shifting $\Phi'_\a$ and $\widehat\Phi'_\ah$ to cancel terms proportional to
$\bar D\T^\a$ and $D\Th^\ah$, the quartic terms in \adsv\ can be reduced to 
\eqn\cancelt{e [ (D\T \g^a \T) (\bar D\Th \g_a \Th) - (D\Th \g^a \Th) (\bar D\T \g_a \T)] .}
Finally, the coefficient $e$ can be scaled to $-{1\over 8}$
by scaling $[\Phi'_\a, \T^\a, \widehat\Phi'_\ah, \Th^\ah]$ appropriately so that the
$AdS_5\times S^5$ action reduces to the flat space action of \twtwo.

\subsec{U(1) generator and d=12 pure spinors}

Expanding in components, the action of \adsr\ is
\eqn\adss{S =r^2 \int d^2 z [(G^{-1}\p G)^J_R (G^{-1}\bar \p G)^R_J}
$$+ \L_R^J (\bar\nabla \tilde\L)^R_J + \widehat \L_R^J (\nabla \tilde{ \widehat\L})_J^R
+ \L^J_R \tilde\L^R_K \widehat \L^K_S \tilde{\widehat \L}{}^S_J -
\tilde\L^R_J \L^J_S \tilde{\widehat \L}{}^S_K \widehat \L^K_R],$$
where the bosonic components are defined by
\eqn\bosdef{\L^J_R = (G^{-1}DG)^J_R {}|_{\k=\bar\k=0}, \quad \tilde\L^R_J= (G^{-1}DG)^R_J {}|_{\k=\bar\k=0},}
$$
\widehat\L^J_R = (G^{-1}\bar DG)^J_R {}|_{\k=\bar\k=0}, \quad \tilde{ \widehat\L}{}^R_J= (G^{-1}\bar DG)^R_J {}|_{\k=\bar\k=0},$$
$$(\bar\nabla \tilde \L)^R_J = \bar\p\tilde\L^R_J + (G^{-1}\bar\p G)^R_S \tilde\L^S_J -
\tilde\L^R_K (G^{-1}\bar\p G)^K_J,$$
$$(\nabla \tilde{\widehat\L})^R_J = \p\tilde{\widehat\L}{}^R_J + (G^{-1}\p G)^R_S \tilde{\widehat\L}{}^S_J -
\tilde{\widehat\L}{}^R_K (G^{-1}\p G)^K_J.$$

In order to construct the U(1) generator needed to define physical states, it is useful to combine the
$SO(4,2)\times SO(6)$ spinors
$(G^{-1} \p G)^J_R$ and $(G^{-1} \p G)_J^R$ into an $SO(10,2)$ spinor $(G^{-1} \p G)^A$ for
$A=1$ to 32 and write
\adss\ in $SO(10,2)$ notation as
\eqn\twa{S =r^2 \int d^2 z d^2 \k ~ \e_{AB} (G^{-1} D G)^A (G^{-1} \bar D G)^B}
\eqn\twb{ = r^2 \int d^2 z [\e_{AB} (G^{-1}\p G)^A (G^{-1}\bar \p G)^B
+\e_{AB} \L^A (\bar\nabla \L)^B +\e_{AB} \widehat \L^A (\nabla \widehat\L)^B}
$$
+ R_{MNPQ} (\L\g^{MN}\L)(\widehat \L \g^{PQ} \widehat\L)],$$
where $A=1$ to 32 are d=12 spinor indices and $M=0$ to 11 are d=12 vector indices, $\e_{AB}$ is the Lorentz-covariant antisymmetric tensor used to raise and lower d=12 spinor indices,
$\g^M$ are the d=12 gamma matrices, $\g^{MN}_{AB} = \g^{MN}_{BA}$ are products
of gamma matrices, 
and
\eqn\fdef{R_{MNPQ} = -\eta_{MP} \eta_{NQ} +\eta_{NP}\eta_{MQ}~~{\rm when}~~0\leq M,N,P,Q\leq 5,}
$$ 
R_{MNPQ} = \eta_{MP} \eta_{NQ} -\eta_{NP}\eta_{MQ}~~{\rm when}~~6\leq M,N,P,Q\leq 11.$$

As in a flat background, a U(1) generator $J$ can be defined which splits $G= G^+ + G^-$ where $G^\pm$ carries
$\pm1$ U(1) charge.
This U(1) generator is constructed by first splitting
the bosonic components $\L^A$ and $\widehat\L^B$ of \twb\ into left and right-moving $d=(10,2)$ pure spinors 
$U^A$ and $\widehat U^A$ satisfying the constraints
\eqn\puretwelve{U^A (\g^{MN})_{AB} U^B =0, \quad \widehat U^A (\g^{MN})_{AB} \widehat U^B =0,}
together with their conjugate momenta
$V_A$ and $\widehat V_A$. 
One then defines the left and right-moving U(1) generators as
\eqn\uone{J = -U^A V_A, \quad
\bar J = -\widehat U^A \widehat V_A,}
so that $U^A$ and $\widehat U^A$ carry $+1$ charge and $V_A$ and $\widehat V_A$ carry $-1$ charge.

Just as a $d=10$ pure spinor parameterizes
${{SO(10)}\over{U(5)}}\times C$ and has 11 independent complex components,
a $d=(10,2)$ pure spinor parameterizes
${{SO(10,2)}\over{U(5,1)}}\times C$ and has 16 independent complex components. So
$J$ splits the 32 components of $\L^A$ into the 16 components of $U^A$ and 
16 components of $V_A$.
To relate $\L^A$ with $U^A$ and $V_A$, introduce a fixed $d=(10,2)$
pure spinor $\bar U_A$ on the patch of pure spinor space where $U^A \bar U_A \neq 0$, and define
\eqn\decompl{\L^A= U^A + \e^{AB} V_B +{1\over {(U\bar U)}} [ {1\over 4}(\g^{MN} U)^A (\bar U\g_{MN}V) +2 U^A (\bar U V) ], }
where the coefficients in \decompl\ have been chosen such that $\bar U \g^{MN} \L = \bar U\g^{MN} U$. Note that
 \decompl\ is invariant under the gauge transformation
\eqn\gaugeV{\d V_A = \Omega^{MN} (\g_{MN} U)_A}
for any $\Omega^{MN}$,
which allows 16 components of $V_A$ to 
be gauged to zero.

When written in terms of $(U^A, V_A)$ and $(\widehat U^A, \widehat V_A)$, the worldsheet action of \twb\ has no terms with positive U(1) charge and the term with zero U(1) charge is
\eqn\adst{S = r^2 \int d^2 z [\e_{AB} (G^{-1}\p G)^A (G^{-1}\bar \p G)^B
+ V_A (\bar\nabla U)^A +\widehat V_A (\nabla \widehat U)^A}
$$
+ R_{MNPQ} (V\g^{MN} U)(\widehat V \g^{PQ} \widehat U)]$$
where $R_{MNPQ}$ is defined in \fdef.
As expected, the term
of zero U(1) charge in \twb\ is independent of the fixed pure spinors $\bar U$ and $\bar{\widehat U}$ and gauge-invariant under \gaugeV.

Under the $SO(4,2)\times SO(6)$ subgroup of $SO(10,2)$, $U^A$ decomposes into
$(\tilde U^R_J, U^J_R)$ and the $d=(10,2)$ pure spinor constraint of \puretwelve\ decomposes
as 
\eqn\dpsc{\tilde U^R_J  U^J_S = {1\over 4}\d^R_S \tilde U^T_J U^J_T, \quad
U^J_R \tilde U^R_K = {1\over 4}\d^J_K  U^L_R \tilde U^R_L,}
$$
\e^{JKLM} \tilde U^R_J \tilde U^S_K = \e^{RSTU}  U^L_T  U^M_U, \quad
\e_{JKLM} U^J_R U^K_S = \e_{RSTU} \tilde U^T_L  \tilde U^U_M.$$
Note that the second line of \dpsc\ is not invariant under the $U(1)$ `bonus' symmetry
which rotates 
\eqn\rotat{U^J_S \to e^{i\phi} U^J_S, \quad \tilde U^S_J \to e^{-i\phi} \tilde U^S_J}
and enlarges the R-symmetry group from $SU(4)$ to $U(4)$. So although the action of \adss\ is invariant
under this $U(1)$ bonus symmetry, $J$ of \uone\ is not invariant which implies that
the action of \adst\ with zero $U(1)$ charge is also not invariant under this bonus symmetry.



\vskip15pt
{\bf Acknowledgements:}

I would like to thank
CNPq grant 300256/94-9
and FAPESP grants 2011/11973-4 and 2014/18634-9 for partial financial support, Nikita Nekrasov and Edward Witten for suggesting to look for an N=1 worldsheet supersymmetric description of the pure spinor formalism, and Andrei Mikhailov, Warren Siegel, Cumrun Vafa and
Pedro Vieira for useful discussions.

\listrefs
\end